\documentclass[prl,superscriptaddress,twocolumn,showpacs,preprintnumbers,amsmath,amssymb]{revtex4}

\usepackage{graphicx}% Include figure files
\usepackage{dcolumn}% Align table columns on decimal point
\usepackage{bm}% bold math
\usepackage{eucal}
\usepackage{ulem}
\usepackage{color}
\begin{document}

\title{Field-induced magnetic phases and electric polarization in LiNiPO$_4$}
\author{T.~B.~S.~Jensen}
\affiliation{Materials Research Department, Ris\o{} DTU, Technical University of
Denmark, DK-4000 Roskilde, Denmark}
\author{N.~B.~Christensen}
\affiliation{Materials Research Department, Ris\o{} DTU, Technical University of
Denmark, DK-4000 Roskilde, Denmark} \affiliation{Laboratory for
Neutron Scattering, ETH Z\"{u}rich and Paul Scherrer Institute,
CH-5232 Villigen, Switzerland}
\author{M.~Kenzelmann}
\affiliation{Laboratory for Neutron Scattering, ETH Z\"{u}rich and
Paul Scherrer Institute, CH-5232 Villigen, Switzerland}
\affiliation{Laboratory for Solid State Physics, ETH Z\"{u}rich,
CH-8093 Z\"{u}rich, Switzerland}
\author{H.~M.~R\o nnow}
\affiliation{Laboratory for Neutron Scattering, ETH Z\"{u}rich and
Paul Scherrer Institute, CH-5232 Villigen, Switzerland}
\affiliation{Laboratory for Quantum Magnetism, Ecole Polytechnique
F\'{e}d\'{e}rale de Lausanne, CH-1015 Lausanne, Switzerland}
\author{C.~Niedermayer}
\affiliation{Laboratory for Neutron Scattering, ETH Z\"{u}rich and
Paul Scherrer Institute, CH-5232 Villigen, Switzerland}
\author{N.~H.~Andersen}
\author{K.~Lefmann}
\affiliation{Materials Research Department, Ris\o{} DTU, Technical University of
Denmark, DK-4000 Roskilde, Denmark}
\author{J.~Schefer}
\affiliation{Laboratory for Neutron Scattering, ETH Z\"{u}rich and
Paul Scherrer Institute, CH-5232 Villigen, Switzerland}
\author{M.~v.~Zimmermann}
\affiliation{Hamburger Synchrotronstrahlungslabor at Deutsches
Elektronen Synchrotron, 22603 Hamburg, Germany}
\author{J.~Li}
\affiliation{Ames Laboratory and Department of Physics and
Astronomy, Iowa State University, Ames, Iowa 50011, USA}
\author{J.~L.~Zarestky }
\affiliation{Ames Laboratory and Department of Physics and
Astronomy, Iowa State University, Ames, Iowa 50011, USA}
\author{D.~Vaknin}
\affiliation{Ames Laboratory and Department of Physics and
Astronomy, Iowa State University, Ames, Iowa 50011, USA}

\date{\today}

\begin{abstract}
Neutron diffraction is used to probe the $(H,T)$ phase diagram of
magneto-electric (ME) LiNiPO$_4$ for magnetic fields along the
$c$-axis. At zero field the Ni spins order in two
antiferromagnetic phases. One has commensurate (C) structures and
general ordering vectors ${\bf{k}}_C = (0,0,0)$,  the other one is
incommensurate (IC) with ${\bf{k}}_{IC} = (0,q,0)$. At low
temperatures the C order collapses above $\mu_0H=12\;\mathrm{T}$ and
adopts an IC structure with modulation vector parallel to ${\bf{k}}_{IC}$. We show
that C order is required for the ME effect and establish how
electric polarization results from a field-induced reduction of
the total magneto-elastic energy.
\end{abstract}
 \maketitle

Materials with both magnetic and electric order as found in
magneto-electric (ME) multiferroics have received growing interest
in recent years \cite{Fiebig,Khomskii06,CheongMostovoy}. It is
expected that the coupling of magnetic and electric order
in multiferroics will be of technological use, but also lead to rich physics with multi-order phase
transitions \cite{Kenzelmann07} and excitations such as
electromagnons \cite{Pimenov,Sushkov}. Often ferroelectric and
magnetic phases have very different ordering
temperatures, suggesting that they are driven by different
microscopic interactions, but for some they coincide and ferroelectricity is generated by
magnetic long-range order \cite{Kimura2003,Lawes05,Kenzelmann05}. In the lithium ortho-phosphates, Li$M$PO$_4$, $M$=Mn, Fe, Co, Ni, a strong ME effect is observed in the antiferromagnetic phases. Mercier \cite{MercierThesis} explained the temperature dependence of the ME coefficients for LiMnPO$_4$, LiCoPO$_4$ and to lesser extent for LiFePO$_4$ with a microscopic model adapted from Cr$_2$O$_3$ \cite{Hornreich67}. However, the ME effect in LiNiPO$_4$ differs from the other lithium phosphates and was not modelled as succesfully. In the present study we determine the field induced magnetic structures in ${\rm
LiNiPO_4}$ for fields $\mathbf{H}||\mathbf{c}$ and correct the existing picture of the zero field structures. Combining symmetry arguments and microscopic calculations similar to \cite{Hornreich67,Sergienko2006} we use the detailed information of the magnetic structures to quantify the ME properties of LiNiPO$_4$ and show how magnetic fields may lead to electric polarization.\par

\begin{figure}[t]
\includegraphics[angle=-90,width=8.5cm]{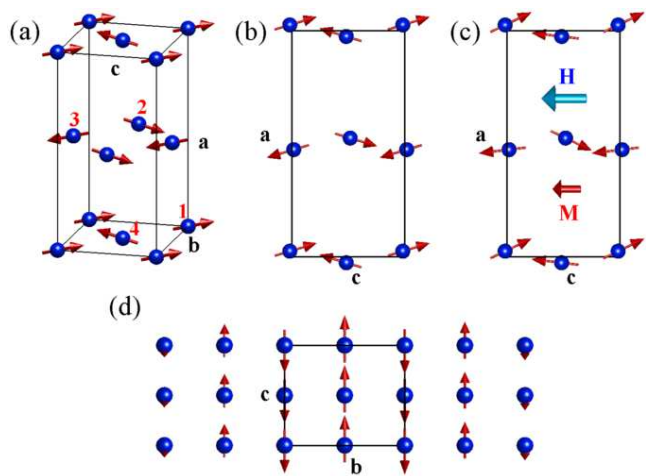}
\caption{(Color online) (a) Ni$^{2+}$ spin configuration of the
zero-field C structure. Ni-positions are labelled {\it i}
according to their positions $\bf{r}_i$ ({\it i} = 1...4) but
shifted $(-0.25,-0.25,0)$ compared to the values given in the text.
(b-c) Projected C structure at zero and finite field
$\mathbf{H}||\mathbf{c}$ seen along the $b$-axis. Spin angles are exaggerated for clarity. The applied field cause the
Ni$^{2+}$ moments to rotate resulting in a magnetization $\mathbf{M}$
of the crystal, as described in the text. (d) Linearly polarized (LP) IC magnetic
structure at zero field seen along the $a$-axis.}
\label{fig:structure_PRL1.eps}
\end{figure}

In LiNiPO$_4$, an electric polarization along the $a$-axis
($c$-axis) is generated when a magnetic field is applied along the
$c$-axis ($a$-axis). This occurs below $T=20.8\;\mathrm{K}$, where
at zero field the material undergoes a first-order transition from
a low temperature commensurate (C) antiferromagnetic phase with a
general ordering vector ${\bf k}_{\rm C} = (0,0,0)$ for each the four Ni-spin sublattices (cf. Fig. \ref{fig:structure_PRL1.eps}a) to an
incommensurate (IC) phase with ${\bf k}_{\rm IC}=(0,q,0)$ and
$0.07 < q < 0.155$ \cite{Vak04}. For fields ${\bf H} || {\bf c}$,
the magnetization measurements provide evidence for several phase
transitions between $\mu_0H=12$ and $22\;\mathrm{T}$ \cite{Khr04}. Also IC magnetic structures 
have been discussed as possible explanation of hysteresis observed in the ME coefficients at high magnetic fields along the $a$-axis \cite{Kornev2000,Chupis2000}.  
\par

To understand the ME effect in
LiNiPO$_4$, we have studied the $(H,T)$ phase diagram and the
magnetic structures for fields ${\bf H} || {\bf c}$ up to
$\mu_0H=14.7\;\mathrm{T}$. We first present the $(H,T)$ phase diagram and
show that off-diagonal single-ion anisotropies and
Dzyaloshinsky-Moriya (DM) interactions allowed by symmetry are consistent with the
observed magnetic structures and lead to staggered magnetic
moments in applied magnetic fields. Then we establish that
electric polarization is only allowed in the field-induced C
structure, but not in the high temperature IC and the zero-field C structures.
Finally we show that the electric polarization is driven by the magnetic symmetry and propose a model accounting
for the temperature dependence of the ME constants.
\par

Measurements were performed on a high-quality $0.4\,\mathrm{g}$
single crystal. Zero field measurements were performed in a closed
cycle cryostat on a four circle goniometer at the TriCS single
crystal diffractometer, using neutron wavelength $\lambda =
1.18$\;\AA\ for the C structure determination at
$T=5\;\mathrm{K}$, and $\lambda = 2.318$\;\AA\; for the IC
structure at $T=21\;\mathrm{K}$. For diffraction measurements on
the triple axis spectrometer RITA-II the sample was mounted in a
$15\;\mathrm{T}$ magnet with the vertical field along the
crystallographic $c$-axis. The $(H,T)$ phase diagram was
determined using neutrons with $\lambda = 4.04$\;\AA{}. The high
field magnetic structure was studied with $\lambda = 2.02$\;\AA\
neutrons.\par

{\it Symmetry properties:} LiNiPO$_4$ crystallizes in the
orthorhombic $Pnma$ (No. 62) crystal structure with lattice
parameters $a = 10.02$ \AA, $b = 5.83$ \AA{} and $c = 4.66$ \AA
\cite{Abrahams93}. The magnetic Ni$^{2+}$ ions with spin $S = 1$
are situated on $4(c)$ sites forming buckled planes perpendicular
to the $a$-axis. The positions of the four
Ni$^{2+}$ in each unit cell are: $\mathbf{r}_1=(0.275,0.25,0.98)$,
$\mathbf{r}_2=(0.775,0.25,0.52)$, $\mathbf{r}_3=(0.725,0.75,0.02)$
and $\mathbf{r}_4=(0.225,0.75,0.48)$, as shown in Fig.
\ref{fig:structure_PRL1.eps}a. The low crystal-field symmetry in
LiNiPO$_4$ leads to a magnetic susceptibility tensor that contains
staggered off-diagonal terms, $\chi_{ac}$ and $\chi_{ca}$. This
allows for a single-ion anisotropy of the following type:
\begin{equation}\label{eq:xz_anisotropy}
\mathcal{H}^{\rm ani}_{zx}=-D_{zx}
(S_1^cS_1^a-S_2^cS_2^a+S_3^cS_3^a-S_4^cS_4^a),
\end{equation}
and two DM interactions:
\begin{equation}\label{eq:DM12}\begin{split}
\mathcal{H}^{\rm DM}_1&=D_1(S_1^cS_2^a-S_2^cS_1^a+S_3^cS_4^a-S_4^cS_3^a), \\
\mathcal{H}^{\rm DM}_2&=-D_2(S_1^cS_4^a-S_4^cS_1^a-S_2^cS_3^a+S_3^cS_2^a).
\end{split}\end{equation}

\begin{figure}[b]
\includegraphics[width=0.7\linewidth,angle=-90]{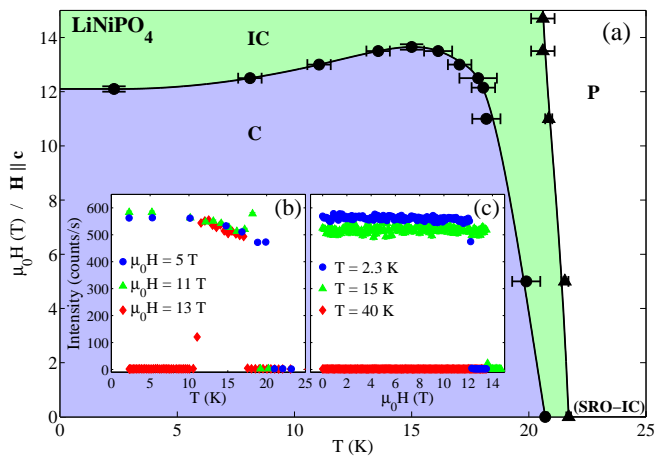}
\caption{(Color online) (a) $(H,T)$ phase diagram of LiNiPO$_4$ for $\mathbf{H}
|| \mathbf{c}$ including a commensurate (C) and an incommensurate
(IC) antiferromagnetic phase, and a paramagnetic (P) phase, which
at zero field supports short range fluctuations up to
$T=40\;\mathrm{K}$ \cite{Vak04}. (b-c) $T$ and $H$ dependence of
the (0,1,0) magnetic Bragg peak intensity at three temperatures
and fields, respectively.} \label{fig:phasediagram}
\end{figure}

\begin{figure}[t]
\includegraphics[width=10.6cm,angle=-90]{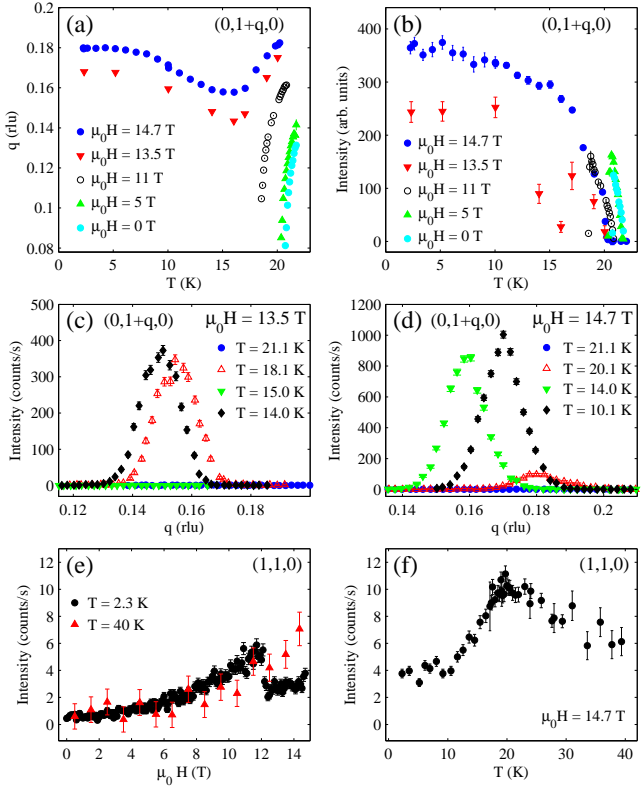}
\caption{(Color online) (a-b) Position and integrated intensity of the
(0,1$+q$,0) IC peak as function of temperature for different
fields ${\bf H}||{\bf c}$. (c-d) Scattering intensity for
wave-vectors $(0,k,0)$ at $\mu_0 H = 13.5\;\mathrm{T}$ and
$14.7\;\mathrm{T}$. At $13.5\;\mathrm{T}$, the IC peak
disappears near $T=15\;\mathrm{K}$ where the system enters the
C phase (cf. Fig.~\ref{fig:phasediagram}a). (e-f) Background
subtracted peak intensities of the (1,1,0) peak as function of
field and temperature.}
\label{fig:figureICdata}
\end{figure}

{\it Phase diagram:} The $(H,T)$ phase diagram for fields
$\mathbf{H} || \mathbf{c}$ up to $\mu_0H = 14.7\;\mathrm{T}$ is shown
in Fig.~\ref{fig:phasediagram}. The C phase is characterized by
commensurate Bragg peaks associated with ordering vector ${\bf
k}_{\rm C}$, such as $(0,1,0)$ whose $T$ and $H$ dependence are
shown in Fig.~\ref{fig:phasediagram}b-c. A sudden drop in
intensity of the $(0,1,0)$ peak indicates the collapse of the C
phase in a first-order phase transition. Between
$T=10\;\mathrm{K}$ and $18\;\mathrm{K}$, the C phase extends to
higher fields, leading to a dome-shaped $(H,T)$ phase diagram. The
C phase is enclosed by an IC phase with a magnetic ordering
wave-vector ${\bf k}_{\rm IC}$, appearing e.g. at $(0,1+q,0)$.
Fig.~\ref{fig:figureICdata}a-d shows the temperature dependence of
$q$ and the intensity of the IC $(0,1+q,0)$ peaks for different
fields.\par

{\it Staggered crystal-fields and DM interactions:} The zero field
C structure belongs to a single irreducible representation of
${\bf k}_{\rm C}$, determined from 112 magnetic peaks at
$5\;\mathrm{K}$. The magnetic moments are nearly parallel to the
$c$-axis with ${\bf m}_{\rm C}=(0.3(1),0,2.2(2))\mu_B$. The
$c$-component, $m_{\rm C}^c$, has a $(+,+,-,-)$ order and the
$a$-component, $m_{\rm C}^a$, a $(+,-,-,+)$ order on the sites
$\mathbf{r}_i$ with increasing $i=1...4$
(Fig.~\ref{fig:structure_PRL1.eps}a-b). Earlier structural
analysis using powder diffraction \cite{Santoro66,Vaknin99} found
$m_{\rm C}^c$, but not the smaller $m_{\rm C}^a$. The presence of
$m_{\rm C}^a$ may be explained by single ion anisotropies and DM
interactions: Inserting $m_{\rm C}^c$ of $(+,+,-,-)$ symmetry into
Eqs. \ref{eq:xz_anisotropy} and \ref{eq:DM12}, we find that
$\mathcal{H}^{\rm ani}_{zx}=-D_{zx} S (S_1^a-S_2^a-S_3^a+S_4^a)$ and $\mathcal{H}^{\rm
DM}_{1,2}=-D_{1,2} S (S_1^a-S_2^a-S_3^a+S_4^a)$, which both favor
that $m_{\rm C}^a$ is of $(+,-,-,+)$ symmetry.\par

{\it Field-induced staggered moments:} Magnetic fields along the
$c$-axis induce antiferromagnetic  $(1,1,0)$ peaks that grow as
($\mu_0H$)$^2$ in the C phase (Fig. \ref{fig:figureICdata}e).
High-energy ($100\;\mathrm{keV}$) X-ray diffraction at the BW5
beam line at HASYLAB, DESY detected no field dependent signal at
$(1,1,0)$, showing that the neutron signal is of magnetic origin.
Structural refinements reveal that the $(1,1,0)$ peak reflects an
additional staggered magnetic $a$-component, $m_{st}^a$, with
$(+,-,+,-)$ symmetry, and an ordered moment increasing linearly
with field to a value of $m_{st} \approx 0.17\mathrm{\mu_B}$ at
$12\;\mathrm{T}$. Assuming that, to first order, the field rotates
the magnetic moments without changing their magnitude, the
magnetic structure (Fig.~\ref{fig:structure_PRL1.eps}c) has a
total ferromagnetic moment of $\sim(0,0,0.6m_{st}^a)$ per unit
cell and the magnetization is $\mathbf{M}\approx (0,0,3.2)$G at 12
T. This is consistent with bulk measurements \cite{Kharchenko03}
showing that the magnetization grows almost linearly with
$\mathbf{H}$ and is $\sim$0.03 G at 0.1 T. At 40 K, where the
system is paramagnetic, there is still field induced (1,1,0)
intensity (Fig. \ref{fig:figureICdata}e). We interpret this as staggered magnetic fields at the Ni position due to
off-diagonal elements of the local susceptibility tensor or the DM
interactions, as previously observed in antiferromagnetic
S=$\frac{1}{2}$ chains \cite{Oshikawa_Affleck,Chen2007}.\par

The IC magnetic order at zero field is a transversely polarized
collinear spin-density wave belonging to a single representation
of ${\bf k}_{\rm IC}$. The structure is shown in
Fig.~\ref{fig:structure_PRL1.eps}d and consists of magnetic
moments ${\bf m}_{\rm IC}=(0.0(2),0.0(1),1.2(2))\mu_B$ that are
ordered with $(+,+,-\beta,-\beta)$ symmetry, where $\beta =
e^{-i\pi q}$ describes the IC modulation along {\bf b}. The IC
component of the high-field structure is found by structural
analysis of $83$ magnetic peaks at $14.7\;\mathrm{T}$ and
$2.3\;\mathrm{K}$. The best fit results in a similar
structure as at zero field, but with increased amplitude ${\bf
m}_{\rm IC}=(0,0,2.0(2))\mu_B$. However, the data does not exclude an elliptically polarized (EP) IC structure with $(+,+,-\beta,-\beta)$ components along $a$ and $c$. An EP IC structure at 14.7 T and 2.3 K is directly supported by analysis of higher order harmonics and indirectly by a mean field calculation predicting a phase boundary between the EP IC structure and a high temperature linearly polarized (LP) IC structure around 15 K at 14.7 T \cite{JensJensenUnpublished}. Coexisting with the IC order
is a field-induced $(1,1,0)$ intensity (Fig.~\ref{fig:figureICdata}e-f) signaling a $(+,-,+,-)$ C-type moment along the $a$-axis of approximately
$0.11\mu_B$.\par

{\it Phenomenology:} The zero field C structure breaks inversion
symmetry, but is invariant under $2_b$ (180$^\circ$ screw axis
along {\bf b}), thus preventing electric polarization
perpendicular to {\bf b}. However, for $\mathbf{H}||\mathbf{c}$
the invariance under $2_b$ is broken and electric polarization is
allowed. The LP IC structures leave at least one point of inversion
invariant and do not allow for electric polarization - even
in the presence of the C staggered moments. This is consistent
with a more formal treatment developed by Harris \cite{Harris07}.\par

\begin{figure}[t!]
\includegraphics[angle=-90,width=8.4cm]{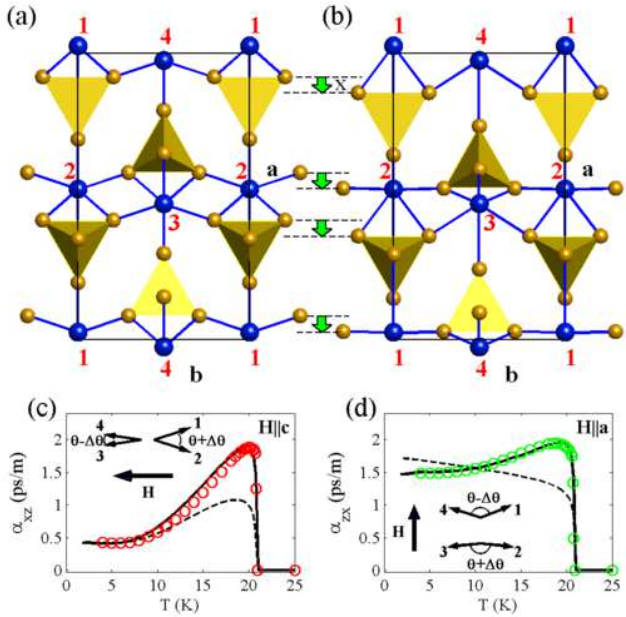}
\caption{(Color online) (a) Positions of Ni (large blue (dark) circles), O (small yellow (bright) circles)
and the PO$_4$-tetrahedra (triangles) in zero field. (b) Same as
(a), but for $\mathbf{H}||\mathbf{c}$. PO$_4$-tetrahedra are assumed to shift downwards with $x$ (arrows), giving an
electric polarization $P_x$ and changing the SE interaction as
explained in the text. (c) Measured (circles) \cite{Vak04} and
calculated ME coefficient $\alpha_{xz}$ for
$\mathbf{H}||\mathbf{c}$, assuming identical spin lengths
(dashed line) and different spin lengths (solid line).
(d) Same as (c) for $\alpha_{zx}$ with $\mathbf{H}||\mathbf{a}$. Insets in (c) and (d) show the assumed angles between
$\mathbf{S}_1$, $\mathbf{S}_2$, $\mathbf{S}_3$ and $\mathbf{S}_4$.}
\label{fig:ME_PRL1.eps}
\end{figure}

{\it Magneto-electric effect:} The main features of the ME effect
in LiNiPO$_4$ can be explained by connections between super-exchange (SE), DM spin interactions and elastic distortions. At zero field $\vert \mathbf{S}_1\vert=\vert \mathbf{S}_2\vert=\vert
\mathbf{S}_3\vert=\vert \mathbf{S}_4\vert= \langle S\rangle$, the
thermal mean value of the spin operator, and the angles between
$\mathbf{S}_1$ and $\mathbf{S}_2$, and between $\mathbf{S}_3$ and
$\mathbf{S}_4$, are identical $\theta_{12}=\theta_{34}=\theta$.
In the C phase a magnetic field $\mathbf{H}||\mathbf{c}$
rotates the spins as shown in Fig. \ref{fig:ME_PRL1.eps}c. Here
$\theta_{12}= \theta+\Delta\theta$ and
$\theta_{34}=\theta-\Delta\theta$, and $\Delta\theta$ is
proportional to the magnetization $\chi_cH_z$, if we assume that $\vert \mathbf{S}_1\vert=\vert \mathbf{S}_2\vert=\vert
\mathbf{S}_3\vert=\vert \mathbf{S}_4\vert$ even at non-zero fields. The SE energy for
$\mathcal{H}^{\rm
SE}_{12,34}=J_{12}\mathbf{S}_1\cdot\mathbf{S}_2+J_{34}\mathbf{S}_3\cdot\mathbf{S}_4$
in this spin configuration is

\begin{equation}
\begin{split}
\mathcal{E}^{\rm SE}_{12,34}=&(J_{12}+J_{34})\langle
S\rangle^2(1-\frac{1}{2}(\theta^2+\Delta\theta^2))\\
&-(J_{12}-J_{34})\langle S\rangle^2\theta\Delta\theta.
\label{eq:SEenergyHc}
\end{split}
\end{equation}

The SE energy (\ref{eq:SEenergyHc}) can be
lowered by a uniform displacement of exchange mediating ions such
as the translation of all PO$_4$-tetrahedra along $\mathbf{a}$ by a
small distance $x$ (Fig. \ref{fig:ME_PRL1.eps}b).
The symmetry of the Ni-O-P-O-Ni exchange paths implies that a
uniform translation of the tetrahedra, leading to an electric
polarization $P_x$ along $\mathbf{a}$, simultaneously increases
$J_{12}$ and reduces $J_{34}$, or vice versa. To first order
$J_{12}=J+\delta$ and $J_{34}=J-\delta$, where $\delta=\lambda_x
x$ for small values of $x$. Introducing an elastic energy
$\epsilon_xx^2$ for the tetrahedra displacements gives a
SE-elastic interaction energy $-2\lambda_x\langle
S\rangle^2\theta\Delta\theta x+\epsilon_xx^2$, which is minimum
for $x = \lambda_x\langle
S\rangle^2\theta\Delta\theta\epsilon_x^{-1}$. Noting that
$P_x\propto x$ and $\Delta\theta\propto\chi_cH_z$ we obtain an
electrical polarization $P_x\propto\epsilon_x^{-1}\langle
S\rangle^2\chi_cH_z$, and thereby a ME coefficient
$\alpha_{xz}\propto\epsilon_x^{-1}\langle S\rangle^2\chi_c$. An equivalent expression for $\alpha_{xz}$ can also be obtained from the DM interaction term $H^{\rm DM}_1$. These ME coefficients are similar to the phenomenological expressions suggested by Rado \cite{Rado1961} but are here established from a microscopical point of view related to \cite{Hornreich67}. Figure
\ref{fig:ME_PRL1.eps}c compares the temperature dependence of the
measured ME coefficient $\alpha_{xz}$ \cite{Vak04} to
$\epsilon_x^{-1}\langle S\rangle^2\chi_c$ (dashed line),
assuming a constant elastic coefficient $\epsilon_x$ and using the
magnetic order parameter $\langle S\rangle$ determined in
\cite{Vaknin99} and the magnetic susceptibility $\chi_c$ from
\cite{Kharchenko03,VakninPrivate}. A more elaborate calculation of $\alpha_{xz}$, assuming the angle difference, $\Delta\theta$,
fixed at the low temperature value, while the
spins have non-identical lengths at finite temperatures,
gives a significantly better agreement with the experimental
data (solid line in Fig. \ref{fig:ME_PRL1.eps}c). Here the expressions for SE and DM are not equivalent and both terms are needed in the best fit to the data.\par

To explain the elastic distortions in the C phase for
$\mathbf{H}||\mathbf{a}$ we first assume, as for
$\mathbf{H}||\mathbf{c}$, that the magnetization of the sample
results from a rotation of the magnetic moments. This way we
obtain the C spin structure sketched in Fig.
\ref{fig:ME_PRL1.eps}d. Using similar arguments, now on the pairs
$(\mathbf{S}_1,\mathbf{S}_4)$ and $(\mathbf{S}_2,\mathbf{S}_3)$,
we find for identical spin lengths a ME coefficient $\alpha_{zx}\propto\epsilon_z^{-1}\langle
S\rangle^2\chi_a$ (dashed line), which is compared to the
measured ME coefficient $\alpha_{zx}$ \cite{Vak04} in
Fig. \ref{fig:ME_PRL1.eps}d. Once again the elaborate calculation (solid line) improves the agreement with the experimental data.

The proposed mechanism for ME distortions is not effective for
$\mathbf{H}||\mathbf{b}$ in the C phase, nor for any field
direction in the LP IC phase. In the former case, all spins will cant
with the same amount in the field direction and have the same
lengths, leading to no energy differences between any pair of
spins and no magneto-electricity. In the LP IC case, the observed
spin structures are superpositions of C and IC components, which
are uncoupled in the energy terms because of translational
symmetry. The symmetries of the C and the IC components
considered separately do not produce the needed energy differences and ME distortions are therefore not induced.\par

{\it Conclusions:} The symmetries of the established magnetic
structures do not
support an electric polarization in the C- and zero field IC phase. Applying a magnetic field along
{\bf c} in the C phase creates a polar axis and allows for
electric polarization. Symmetry analysis show that
electric polarization is possible in the C phase structure, but
not in the LP IC phase. A microscopic model explains the
temperature dependence of the ME constants, providing
evidence that the electric polarization in LiNiPO$_4$ results from field induced changes in the magnetic structure.\par

Jens Jensen is greatly acknowledge for illuminating discussions. Work was supported by the Danish Agency for Science,
Technology and Innovation under DANSCATT and by the Swiss NSF via
contract PP002-102831. This manuscript has been authored,  
in whole or in part, under Contract No. DE-AC02-07CH11358 with the  
U.S. Department of Energy.
Experiments were performed at the SINQ neutron spallation source
at the Paul Scherrer Institute, Switzerland.


\begin{thebibliography}{01}

\bibitem{Fiebig} M. Fiebig, J.\ Phys. D:\ Appl.\ Phys.\ {\bf 38},
R123 (2005)
\bibitem{Khomskii06} D.I. Khomskii, J.\ Magn.\ Magn. \ Mat. \ {\bf
306}, 1-8 (2006)
\bibitem{CheongMostovoy} S.~W. Cheong and M. Mostovoy, Nature Materials {\bf 6}, 13
(2007).
\bibitem{Kenzelmann07} M. Kenzelmann, G. Lawes, A.~B. Harris, G. Gasparovic, C. Broholm, A.~P. Ramirez, G.~A. Jorge, M. Jaime, S. Park, Q. Huang, A.~Ya. Shapiro and L.~A. Demianets, Phys. Rev. Lett. {\bf 98}, 267205 (2007).
\bibitem{Pimenov}A. Pimenov, A.~A. Mukhin, V.~Yu. Ivanov, V.~D. Travkin, A.~M. Balbashov and A. Loidl, Nature Physics {\bf 2} 97
(2006).
\bibitem{Sushkov}A.~B. Sushkov, R.~Vald\'es Aguilar, S. Park, S-W. Cheong and H.~D. Drew, Phys. Rev. Lett. {\bf
98}, 027202 (2007).
\bibitem{Kimura2003} T. Kimura, T. Goto, H. Shintani, K. Ishizaka, T. Arima and Y. Tokura, Nature {\bf 426}, 55 (2003)
\bibitem{Lawes05} G. Lawes, A.~B. Harris, T. Kimura, N. Rogado, R.~J. Cava, A. Aharony, O. Entin-Wohlman, T. Yildirim, M. Kenzelmann, C. Broholm and A.~P. Ramirez,
Phys.\ Rev.\ Lett.\ {\bf95}, 087205 (2005).
\bibitem{Kenzelmann05} M. Kenzelmann, A.~B. Harris, S. Jonas, C. Broholm, J. Schefer, S.~B. Kim, C.~L. Zhang, S.-W. Cheong, O.~P. Vajk and J.~W. Lynn,
Phys.\ Rev.\ Lett.\ {\bf95}, 087206 (2005).
\bibitem{MercierThesis} M. Mercier, Ph.D. thesis, Universit\'e de Grenoble (1969).
\bibitem{Hornreich67} R. Hornreich and S. Shtrikman,
Phys.\ Rev.\ {\bf161}, 506 (1967).
\bibitem{Sergienko2006} I. A. Sergienko and E. Dagotto, Phys. Rev. B. {\bf73}, 094434 (2006).
\bibitem{Vak04} D. Vaknin, J.~L. Zarestky, J.-P. Rivera, and H. Schmid,
Phys.\ Rev.\ Lett.\ {\bf92}, 207201 (2004).
\bibitem{Khr04} V.~M. Khrustalyov, V. N. Savitsky and N. F. Kharchenko,
Czech.\ J.\ Phys.\ {\bf54}, D27 (2004).
\bibitem{Kornev2000} I. Kornev, M. Bichurin, J.-P. Rivera, S. Gentil, H. Schmid, A.~G.~M. Jansen and P. Wyder, Phys.\ Rev.\ B\ {\bf62}, 12247 (2000).
\bibitem{Chupis2000} I. E. Chupis, Low Temp.\ Phys.\ {\bf26}, 419 (2000).
\bibitem{Abrahams93} I. Abrahams and K. S. Easson,
Acta.\ Cryst.\ {\bf C49}, 925 (1993).
\bibitem{Santoro66} R. P. Santoro, D. J. Segal and R. E. Newnham,
J.\ Phys.\ Chem.\ Solids.\ {\bf27}, 1192 (1966).
\bibitem{Vaknin99} D. Vaknin, J.~L. Zarestky, J.~E. Ostenson, B.~C. Chakoumakos, A. Go\~ni, P.~J. Pagliuso, T. Rojo and G.~E. Barberis,
Phys.\ Rev.\ B\ {\bf60}, 1100 (1999).
\bibitem{Kharchenko03} Yu.~N. Kharchenko, N.~F. Kharcheno, M. Baran and R. Szymczak,
Low Temp.\ Phys.\ {\bf29}, 579 (2003).
\bibitem{Oshikawa_Affleck} M. Oshikawa and I. Affleck, Phys. Rev. Lett. {\bf79}, 2883 (1997).
\bibitem{Chen2007} Y. Chen, M.~B. Stone, M. Kenzelmann, C.~D. Batista, D.~H. Reich and C. Broholm, Phys. Rev. B. {\bf75}, 214409 (2007).
\bibitem{Harris07} A.~B. Harris, Phys.\ Rev.\ B\ {\bf 76}, 054447 (2007).
\bibitem{Rado1961} G.~T. Rado, Phys.\ Rev.\ Lett.\ {\bf 6}, 609 (1961).
\bibitem{VakninPrivate} D. Vaknin, private communication.
\bibitem{JensJensenUnpublished} J. Jensen, private communication.



\end{thebibliography}
\end{document}